\documentclass{article}
\usepackage{epsf,mathrsfs}

\def\vec#1{\mbox{\boldmath $#1$}}
\def\omit#1{_{\!\rlap{$\scriptscriptstyle \backslash$}
{\scriptscriptstyle #1}}}
\def\address#1{\\ {\small\em #1}}

\title{\bf Neutrino scattering in strong magnetic
fields\thanks{Plenary talk given by Palash B. Pal at the COSMO-99
conference held at the Abdus Salam ICTP, Trieste, Italy, during
29/9/1999 to 2/10/1999}}

\author{Kaushik Bhattacharya and Palash B. Pal
\address{Saha Institute of Nuclear Physics, Calcutta 700064, India}}

\date{}

\begin{document}

\maketitle 

\section{Motivation}
Neutrino interactions in strong magnetic fields have gained a lot of
attention because of the problem of neutrino emission from a
proto-neutron star. It has been argued that a small asymmetry in the
neutrino emission can explain the high peculiar velocities of
pulsars. Based on earlier work\cite{magdisp} on neutrino dispersion
relations in an external magnetic field, it has been shown\cite{KuSe}
that neutrino oscillations can probably account for the pulsar
kicks. On the other hand, there is a parallel stream of argument with
the idea that neutrino opacities may change appreciably in magnetic
fields. Asymmetric opacities can presumably account for the kicks. The
opacity has recently been calculated by Roulet\cite{roulet}, assuming
the matrix element for the interactions remain unchanged in a magnetic
field. He found that the opacities do change appreciably compared to
the no-field case. However, he did not find any asymmetry vis-a-vis
the direction of the magnetic field. The aim of the present work is to
redo these calculations from first principles and show that the
relevant cross-section is indeed asymmetric.

\section{Solutions of the Dirac equation in a uniform magnetic
field}
We take the magnetic field $B$ in the $z$-direction, and choose
\begin{eqnarray}
A_0 = A_y = A_z = 0 \,, \qquad A_x = -By\,.
\label{A}
\end{eqnarray}
The Dirac equation in this field can be exactly solved. The energy
eigenvalues are given by
\begin{eqnarray}
E_n^2 = m^2 + p_z^2 + 2neB \,,
\label{En}
\end{eqnarray}
where $n$ is a non-negative integer signifying the Landau level. All
$n>0$ levels are doubly degenerate, whereas the $n=0$ level is
non-degenerate. The eigenstates are of the form 
\begin{eqnarray}
e^{-ip\cdot X {\omit y}} U_s (y,n,\vec p \omit y) \,,
\end{eqnarray}
where the notation $p\cdot X {\omit y}$ stands for the dot product,
setting the $y$-components equal to zero. $U_s$ is the spinor, given
by 
\begin{eqnarray}
U_+  = \left( \begin{array}{c} 
I_n(\xi) \\[2ex] 0 \\[2ex] 
{\textstyle p_z \over \textstyle E_n+m} I_n(\xi) \\[2ex]
{\textstyle M_n \over \textstyle 
E_n+m} I_{n-1} (\xi) 
\end{array} \right) \,, \quad 
U_- = \left( \begin{array}{c} 
0 \\[2ex] I_{n-1} (\xi) \\[2ex]
{\textstyle M_n \over \textstyle E_n+m} I_n(\xi) \\[2ex]
-\,{\textstyle p_z \over \textstyle 
E_n+m} I_{n-1}(\xi) 
\end{array} \right) \,.
\label{Usoln}
\end{eqnarray}
We have used the shorthand $M_n = \sqrt{2neB}$, and a dimensionless
variable 
\begin{eqnarray}
\xi = \sqrt{eB} \left( y + {p_x \over eB} \right) \,.
\label{xi}
\end{eqnarray}
The function $I_n(\xi)$ appearing in Eq.\ (\ref{Usoln}) is given by
\begin{eqnarray}
I_n(\xi) = N_n e^{-\xi^2/2} H_n(\xi) \,,
\label{In}
\end{eqnarray}
where $H_n$ are Hermite polynomials, and $N_n$ is a normalization
which can be chosen arbitrarily. We choose
\begin{eqnarray}
N_n = 
\left( {\sqrt{eB} \over n! \, 2^n \sqrt{\pi}} \right)^{1/2} \,.
\label{Nn}
\end{eqnarray}
For the sake of consistency, we will define $I_{-1}=0$ so that the
solution $U_-$ vanishes for the zeroth Landau level.  From the spinor
solutions, the spin sum can be calculated:
\begin{eqnarray}
&& \hspace{-1.5cm} P_U (y,y_\star ,n,\vec p\omit y) \equiv
\sum_s U_s (y,n,\vec p\omit y) \overline U_s (y_\star ,n,\vec p\omit
y) \nonumber\\*  
& = &
{1\over 2(E_n+m)} 
\bigg[ \left\{ m(1+\sigma_z) +
\rlap/p_\parallel - 
\widetilde{\rlap/p}_\parallel \gamma_5 \right\} I_n(\xi)
I_n(\xi_\star) \nonumber\\*
&+& \left\{ m(1-\sigma_z) + \rlap/p_\parallel +
\widetilde{\rlap/p}_\parallel \gamma_5 \right\} I_{n-1}(\xi)
I_{n-1} (\xi_\star) \nonumber\\*
&+& M_n \Big\{ (\gamma_1 + i\gamma_2) I_n(\xi) I_{n-1}(\xi_\star) 
+ (\gamma_1 - i\gamma_2) I_{n-1}(\xi) I_n(\xi_\star) \Big\} \Bigg] \,,
\label{PU}
\end{eqnarray}
using the notations $\rlap/p_\parallel=\gamma_\mu p^\mu_\parallel$, 
$\widetilde{\rlap/p}_\parallel=\gamma_\mu \widetilde p^\mu_\parallel$,
where 
\begin{eqnarray}
p_\parallel^\mu = (p_0, 0, 0, p_z) \,, \qquad
\widetilde p_\parallel^\mu = (p_z, 0, 0, p_0) \,. 
\end{eqnarray}
%

\section{The fermion field operator}
The fermion field operator can be written as 
\begin{eqnarray}
\psi(X)  &=&  
\sum_{s=\pm} \sum_{n=0}^\infty \int {dp_x \, dp_z \over
2\pi} \sqrt {E_n+m \over 2E_n} \nonumber\\* & \times &
\Bigg[ f_s (n,\vec p\omit y) e^{-ip\cdot X {\omit y}} U_s (y,n,\vec
p\omit y) 
+ \widehat f_s^\dagger (n,\vec p\omit y) e^{ip\cdot X {\omit y}} V_s
(y,n,\vec p\omit y) \Bigg] \,.
\label{psi}
\end{eqnarray}
where $f_s(n,\vec p\omit y)$is annihilation operator for the fermion,
and $\widehat f_s^\dagger(n,\vec p\omit y)$ is the creation operator
for the antifermion in the $n$-th Landau level with given values of
$p_x$ and $p_z$. The creation and annihilation operator satisfy the
anticommutation relations
\begin{eqnarray}
\left[ f_s (n,\vec p\omit y), f_{s'}^\dagger (n',\vec p'\omit y)
\right]_+ = 
\delta_{ss'} \delta_{nn'} \delta(p_x-p'_x) \delta (p_z - p'_z) \,,
\label{freln}
\end{eqnarray}
etc. The prefactor appearing in Eq.\ (\ref{psi}) is determined from
the fact that the field operator should satisfy the anticommutation
rule 
\begin{eqnarray}
\left[ \psi(X), \psi^\dagger(X_\star) \right]_+ = \delta^3 (\vec X - \vec
X_\star) 
\label{anticomm}
\end{eqnarray}
for $X^0=X_\star^0$. 
The one-fermion states are defined by 
\begin{eqnarray}
\left| n,\vec p\omit y \right> = {2\pi \over \sqrt{L_x L_z}} \;  f^\dagger (n,\vec p\omit y) \left|
0 \right> \,,
\end{eqnarray}
the normalization constant chosen so that these states are orthonormal
in a box with sides $L_x$, $L_y$ and $L_z$. Then
\begin{eqnarray}
\Big< n,\vec p\omit y \Big|\; \overline \psi_U(X) 
&=& \sqrt {E_n+m \over 2E_n L_xL_z}
e^{ip\cdot X \omit y} \overline U_s (y,n,\vec p\omit y) \big< 0
\big| \,,
\label{brapsi}
\end{eqnarray}
where $\psi_U$ denotes the term in Eq.\ (\ref{psi}) that contains the
$U$-spinors. 

\section{Inverse beta-decay}
We now apply the above formalism for the inverse beta-decay process
\begin{eqnarray}
\nu_e(\vec k) + n(\vec P) \to p(\vec P') + e(\vec p'\omit y, n') \,.
\label{invbeta}
\end{eqnarray}
Assume $eB\ll m_p^2$, so that the magnetic field effects on the proton
and neutron spinors can be ignored.

The interaction Lagrangian is
\begin{eqnarray}
{\mathscr L}_{\rm int} = \sqrt 2 \,G_\beta \left[ \overline
\psi_{(e)} \gamma^\mu L \psi_{(\nu_e)} \right] \; 
\left[ \overline
\psi_{(p)} \gamma_\mu (g_V+g_A \gamma_5) \psi_{(n)} \right] \,,
\end{eqnarray}
where $G_\beta=G_F\cos\theta_c$, $g_V=1$ and $g_A=1.26$. For the
$S$-matrix element, this gives 
\begin{eqnarray}
S_{fi} &=& \sqrt 2 \, G_\beta \int d^4X 
\left< e(\vec p'\omit y, n') \left| \overline
\psi_{(e)} \gamma^\mu L \psi_{(\nu_e)} 
\right| \nu_e (\vec k) \right> \nonumber\\*
&& \times \left< p(P') \left| \overline
\psi_{(p)} \gamma_\mu (g_V+g_A \gamma_5) \psi_{(n)}  \right| n(P)
\right> \,. 
\label{Sfi1}
\end{eqnarray}
The hadronic part of the matrix element is obtained with usual rules,
which gives
\begin{eqnarray}
{e^{i(P'-P)\cdot X} \over \sqrt{2P_0 V}
\sqrt{2P_0' V}} \; 
\left[ \overline u_{(p)}(\vec P') \gamma_\mu (g_V+g_A \gamma_5)
u_{(n)}(\vec P) \right]  \,,
\end{eqnarray}
For the leptonic part, we need magnetic spinors for the
electron. Using Eq.\ (\ref{brapsi}), we obtain this matrix element to
be
\begin{eqnarray}
{e^{-ik\cdot X + ip'\cdot X\omit y} \over \sqrt{2\omega V}}
\sqrt{E_{n'}+m \over 2E_{n'}L_xL_z} 
\Big[ \overline U_{(e)}(y,n',\vec p'\omit y) \gamma^\mu L
u_{(\nu_e)}(\vec k) \Big] .
\end{eqnarray}
Putting these back into Eq.\ (\ref{Sfi1}) and performing the
integrations over all co-ordinates except $y$, we obtain
\begin{eqnarray}
S_{fi} = (2\pi)^3 \delta^3 \omit y (P+k-P'-p')
\left[E_{n'}+m \over 2\omega V \; 2P_0V \; 2P_0'V
2E_{n'}L_xL_z \right]^{1/2} {\mathscr M}_{fi} \,,
\label{Sfi2}
\end{eqnarray}
where
\begin{eqnarray}
{\mathscr M}_{fi} &=& \surd 2 G_\beta
\Big[ \overline u_{(p)}(\vec P') \gamma^\mu (g_V+g_A\gamma_5)
u_{(n)}(\vec P) \Big] \nonumber\\*   
&\times & \int dy \; e^{iq_yy}
\Big[ \overline U_{(e)} (y,n',\vec p'\omit y) \gamma_\mu L
u_{(\nu_e)}(\vec k) \Big] \,,
\end{eqnarray}
using the shorthand
\begin{eqnarray}
q_y = P_y+k_y-P'_y \,.
\end{eqnarray}
Note that Eq.\ (\ref{Sfi2}) does not have a $\delta$-function
corresponding to the conservation of $y$-component of momentum,
because the latter is not a good quantum number in this problem.
The transition rate can now be written as 
\begin{eqnarray}
|S_{fi}|^2/T &=& {1\over 16} (2\pi)^3 \delta^3 \omit y (P+k-P'-p')
{E_{n'}+m \over V^3 \omega P_0 P_0' E_{n'}}
\Big| {\mathscr M}_{fi} \Big|^2 \,.
\end{eqnarray}

Putting in now the differential phase space for final particles, which
is
\begin{eqnarray}
{L_x\over 2\pi} \, dp'_x \; {L_z\over 2\pi} \, dp'_z \;
{V\over (2\pi)^3} \, d^3P' 
\label{drho}
\end{eqnarray}
where $V=L_xL_yL_z$, we can write the differential cross section as
\begin{eqnarray}
d\sigma 
= {1\over 64\pi^2} \delta^3 \omit y (P+k-P'-p') \;
{E_{n'}+m \over \omega P_0P_0' E_{n'}} \;
\Big| {\mathscr M}_{fi} \Big|^2 {L_xL_z\over V} 
\, dp'_x \, dp'_z \; d^3P' \,.
\label{dsigma}
\end{eqnarray}
The square of the matrix element, averaged over initial neutron spin,
is given by
\begin{eqnarray}
\Big| {\mathscr M}_{fi} \Big|^2 = G_\beta^2 \ell_{\mu\nu} H^{\mu\nu} \,.
\end{eqnarray}
Here, $H^{\mu\nu}$ is the hadronic part, which is
\begin{eqnarray}
H^{\mu\nu} &=& 4(g_V^2 + g_A^2) (P^\mu P'^\nu + P^\nu P'^\mu -
g^{\mu\nu} P \cdot P') \nonumber\\*
&+& 4 (g_V^2-g_A^2) m_n m_p g^{\mu\nu} + 8g_V
g_A i \varepsilon^{\mu\nu\lambda\rho} P'_\lambda P_\rho \,.
\end{eqnarray}
The leptonic part $\ell_{\mu\nu}$ contains magnetic spinors:
\begin{eqnarray}
\ell_{\mu\nu} = \int dy \int dy_\star \; e^{iq_y(y-y_\star)} \; {\rm Tr}
\Big[ P_U (y_\star, y, n', \vec p' \omit y) \gamma_\mu \rlap/k
\gamma_\nu L \Big] \,.
\end{eqnarray}
The integrations over $y$ and $y_\star$ can be exactly
performed\cite{GradRyzh}, yielding
\begin{eqnarray}
\ell_{\mu\nu} = {2\pi\over eB} \; {1 \over (E_{n'}+m)} (\Lambda_\mu
k_\nu + \Lambda_\nu k_\mu - k \cdot \Lambda g_{\mu\nu} - i
\varepsilon_{\mu\nu\alpha\beta} \Lambda^\alpha k^\beta) \,,
\end{eqnarray}
where
\begin{eqnarray}
\Lambda^\alpha &=& \left[I_n 
\left({q_y \over \sqrt{eB}} \right) \right]^2 
(p'^\alpha_\parallel - \widetilde p'^\alpha_\parallel) 
+ \left[ I_{n-1} \left({q_y \over \sqrt{eB}} \right) \right]^2 
(p'^\alpha_\parallel + \widetilde p'^\alpha_\parallel) \nonumber\\* && +  
2 M_n g_2^\alpha I_n
\left({q_y \over \sqrt{eB}} \right) I_{n-1}
\left({q_y \over \sqrt{eB}} \right) \,.
\label{Lambda}
\end{eqnarray}

We calculate the cross section in the rest frame of the
neutron. The axes are chosen such that incoming neutrino is in the
$x$-$z$ plane. Also, we assume that in the energy regime of interest,
$|\vec P'| \ll m_p$ so that the proton is non-relativistic.
Then we obtain, for $n'\neq 0$, 
\begin{eqnarray}
\Big| {\mathscr M}_{fi} \Big|^2 = {16\pi G_\beta^2 \over eB} \;
{m_nm_p \over E_{n'}+m} \Big[ (g_V^2+3g_A^2) \omega \Lambda_0
+ (g_V^2-g_A^2) k_z \Lambda_z \Big] \,.
\label{Mfisq}
\end{eqnarray}
We now put it into Eq.\ (\ref{dsigma}).  Integrations over $P'_x$ and
$P'_z$ get rid of the $\delta$-functions.  Integration over $p'_x$
gives a factor $L_yeB$.
\begin{eqnarray}
d\sigma 
= {G_\beta^2\over 4\pi} {\delta (Q+\omega-E_{n'})
\over \omega E_{n'}} \;
\Big[ (g_V^2+3g_A^2) \omega \Lambda_0
+ (g_V^2-g_A^2) k_z \Lambda_z \Big] \; dP'_y dp'_z \,,
\label{dsigma2}
\end{eqnarray}
where $Q=m_n-m_p$.  Integration over $P'_y$ can now be performed using
the orthogonality of the Hermite polynomials. This ensures that the
$M_n$-term of Eq.\ (\ref{Lambda}) vanishes. The other two give
\begin{eqnarray}
d\sigma_{n'>0} 
= {eBG_\beta^2\over 2\pi} \delta (Q+\omega-E_{n'})
\Bigg[ (g_V^2+3g_A^2) 
+ (g_V^2-g_A^2) {k_z p'_z  \over \omega E_{n'}} \Bigg]  dp'_z \,.
\end{eqnarray}
The argument of the remaining $\delta$-function vanishes for two
different values of $p'_z$ which are equal and opposite. The term
proportional to $g_V^2-g_A^2$ then cancels out and we obtain
\begin{eqnarray}
\sigma_{n'>0}
&=& {eBG_\beta^2\over \pi} 
(g_V^2+3g_A^2) {Q+\omega \over \sqrt{(Q+\omega)^2 - m^2 - M_{n'}^2}}
\,.
\label{neq0}
\end{eqnarray}

For $n'=0$, however, even the second term of Eq.\ (\ref{Lambda})
vanishes, so we need to redo everything beginning from Eq.\
(\ref{Mfisq}). The final result is:
\begin{eqnarray}
\sigma_0
&=& {eBG_\beta^2\over 2\pi} 
\left[ (g_V^2+3g_A^2)
- (g_V^2-g_A^2) {k_z \over \omega} \right] {Q+\omega  \over
\sqrt{(Q+\omega)^2 - m^2}} \,.
\label{sigma0}
\end{eqnarray}
%

\begin{figure}[bhtp]
\centerline{\epsfxsize=.7\textwidth\epsfbox{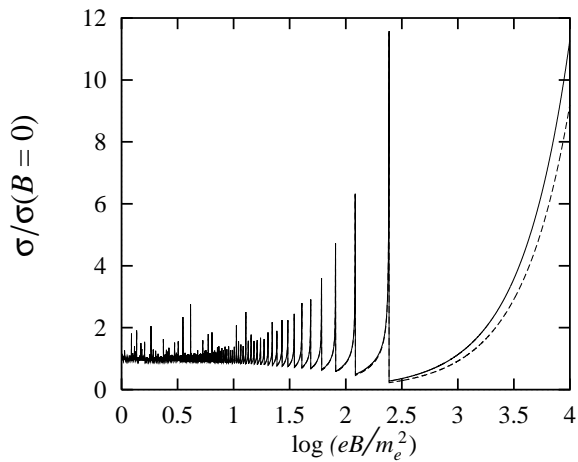}}
\caption[]{Total cross section as a function of the magnetic field,
normalized to the cross section in the field-free case. The initial
neutrino energy is 10\,MeV. The solid and the dashed lines are for the
initial neutrino momentum parallel and antiparallel to the magnetic
field.}\label{f:sigma}
\end{figure}
For any given energy of the initial states, there is a maximum
possible value for $n'$, dictated by Eq.\ (\ref{En}). The total cross
section is obtained by performing the sum over all possible $n'$. We
can then summarize our results as follows:
\begin{list}{$\bullet$}{\setlength{\rightmargin}{\leftmargin}\itemsep=0pt}

\item $\sigma_0$ depends on the direction of neutrino momentum because
of the term involving $k_z$.

\item Total $\sigma$ always contains $\sigma_0$, and is hence asymmetric.

\item The fractional amount of asymmetry depends on $n'_{\rm max}$. It
is maximum if $n'_{\rm max}=0$, and decreases at smaller values of $B$
as more Landau levels contribute.

\end{list}
In Fig.~\ref{f:sigma}, we show the cross section as a function of $B$
for neutrinos incident along and opposite to the direction of the
magnetic field. For small values of $B$, the distinction between the
two directions is not seen at the scale of the plot. However, when $B$
is so large that $n'_{\rm max}=0$, the asymmetry is large. It is, in
fact, as large as 18\% between these two extreme directions. It seems
highly plausible that such a large asymmetry in the cross section
would produce enough momentum asymmetry to explain the pulsar
kicks. These calculations were not finished at the time of the
conference and therefore could not be presented\cite{BhPa}.

There is however another important issue to be settled before the
momentum asymmetry is calculated. Neutrinos undergo many other
reactions inside a dense star. The inverse beta decay has the largest
cross section for the $B=0$ case of course, but there is no guarantee
that other cross sections are not larger in strong magnetic
fields. Our preliminary calculations for the neutrino-electron elastic
scattering show that it can increase by several orders of magnitude in
a magnetic field. If that is the case, then the opacities have to be
calculated taking these processes into account.

\end{document}